\documentclass[numberedappendix,twocolappendix,iop,apj]{emulateapj}
\usepackage{psfig,amsfonts,amsmath,graphicx,natbib,nicefrac,longtable,graphicx
}

\usepackage[backref,breaklinks,colorlinks,citecolor=blue,urlcolor=blue]{hyperref}

\usepackage{booktabs}

\newcommand{\Swift}{\textit{Swift}}
\newcommand{\Konus}{\textit{Konus-Wind}}
\newcommand{\Fermi}{\textit{Fermi}}

\newcommand{\EKiso}{\ensuremath{E_{\rm K,iso}}}

\newcommand{\epse}{\ensuremath{\epsilon_{\rm e}}}
\newcommand{\epsb}{\ensuremath{\epsilon_{\rm B}}}

\newcommand{\Astar}{\ensuremath{A_{*}}}
\newcommand{\tdec}{\ensuremath{t_{\rm dec}}}
\newcommand{\tjet}{\ensuremath{t_{\rm jet}}}

\newcommand{\nua}{\ensuremath{\nu_{\rm a}}}

\newcommand{\numax}{\ensuremath{\nu_{\rm m}}}
\newcommand{\nuc}{\ensuremath{\nu_{\rm c}}}
\newcommand{\nuaf}{\ensuremath{\nu_{\rm a,f}}}
\newcommand{\numf}{\ensuremath{\nu_{\rm m,f}}}
\newcommand{\nucf}{\ensuremath{\nu_{\rm c,f}}}

\newcommand{\nuar}{\ensuremath{\nu_{\rm a,r}}}
\newcommand{\numr}{\ensuremath{\nu_{\rm m,r}}}
\newcommand{\nucr}{\ensuremath{\nu_{\rm c,r}}}
\newcommand{\fnumr}{\ensuremath{F_{\nu,\rm m,r}}}

\newcommand{\nuopt}{\ensuremath{\nu_{\rm opt}}}
\newcommand{\nux}{\ensuremath{\nu_{\rm X}}}

\newcommand{\fnumax}{\ensuremath{F_{\nu,\rm m}}}

\shorttitle{First detection of mm-band polarization in a GRB}
\shortauthors{Laskar et al.}

\def\bath{1}
\def\cfa{4}
\def\northwestern{2}
\def\lvjm{7}
\def\cssac{5}
\def\mpe{6}
\def\ouj{3}
\def\cierafellow{*}

\begin{document} 
\submitted{Published in ApJL}

\title{ALMA Detection of a Linearly Polarized Reverse Shock in GRB\,190114C}
\author{Tanmoy~Laskar\altaffilmark{\bath}}
\author{Kate~D.~Alexander\altaffilmark{\northwestern\cierafellow}}
\author{Ramandeep~Gill\altaffilmark{\ouj}}
\author{Jonathan~Granot\altaffilmark{\ouj}}
\author{Edo~Berger\altaffilmark{\cfa}}
\author{C.~G.~Mundell\altaffilmark{\bath}}
\author{Rodolfo~Barniol Duran\altaffilmark{\cssac}}
\author{J.~Bolmer\altaffilmark{\mpe}}
\author{Paul~Duffell\altaffilmark{\cfa}}
\author{Hendrik~van~Eerten\altaffilmark{\bath}}
\author{Wen-fai~Fong\altaffilmark{\northwestern}}
\author{Shiho~Kobayashi\altaffilmark{\lvjm}}
\author{Raffaella~Margutti\altaffilmark{\northwestern}}
\author{Patricia~Schady\altaffilmark{\bath}}

\affil{\altaffilmark{\bath}Department of Physics, University of Bath, Claverton Down, Bath BA2 
7AY, United Kingdom}
\affil{\altaffilmark{\northwestern}Center for Interdisciplinary Exploration and Research in 
Astrophysics (CIERA) and Department of Physics and Astronomy, Northwestern University, Evanston, 
IL 60208, USA}
\affil{\altaffilmark{\ouj}Department of Natural Sciences, The Open University of Israel, PO Box 
808, Ra'anana 43537, Israel}
\affil{\altaffilmark{\cfa}Center for Astrophysics, Harvard \& Smithsonian, 60 Garden Street, 
Cambridge, MA 02138, USA}
\affil{\altaffilmark{\cssac}Department of Physics and Astronomy, California State University, 
Sacramento, 6000 J Street, Sacramento, CA 95819, USA}
\affil{\altaffilmark{\mpe}Max-Planck-Institut f\"ur extraterrestrische Physik, 
Giessenbachstra{\ss}e, D-85748 Garching, Germany}
\affil{\altaffilmark{\lvjm}Astrophysics Research Institute, Liverpool John Moores University, 
IC2, Liverpool Science Park, 146 Brownlow Hill, Liverpool L3 5RF, United Kingdom}

\altaffiltext{\cierafellow}{Einstein Fellow}

\begin{abstract}
We present Atacama Large Millimeter/submillimeter Array 
97.5~GHz total intensity and linear polarization observations of the mm-band 
afterglow of GRB~190114C spanning 2.2--5.2 hours after the burst.  
We detect linear polarization at the $\approx 5\,\sigma$ level, decreasing from 
$\Pi=(0.87\pm0.13)\%$ to $(0.60\pm0.19)\%$, and evolving in polarization position angle from 
$(10\pm5)^\circ$ to $(-44\pm12)^\circ$ during the course of the observations. 
This represents the first detection {and measurement of the temporal evolution} of polarized 
radio/millimeter emission in a $\gamma$-ray burst.
We show that the optical and X-ray observations between $0.03$~days and $\sim0.3$~days are 
consistent with a fast-cooling forward shock expanding into a wind environment. However, the 
optical observations at $\lesssim0.03$~days, as well as the radio and millimeter observations, arise 
from a separate component, which we interpret as emission from the reverse-shocked ejecta. 
Using the measured linear polarization, we constrain the coherence scale of tangled magnetic fields 
in the ejecta to an angular size of $\theta_{\rm B} \approx10^{-3}$~radian, while 
the rotation of the polarization angle rules out the presence of large-scale, ordered axisymmetric 
magnetic fields, and in particular a large scale toroidal field, in the jet.

\end{abstract}


\section{Introduction}
The interaction of gamma-ray burst (GRB) jets with their ambient medium generates two shocks: (i) a 
relativistic forward shock (FS) in the ambient medium that powers the long-lasting X-ray to radio 
afterglow radiation, and (ii) a short-lived reverse shock (RS) propagating into, and decelerating, 
the jet \citep{spn98,zk05}. 
Whereas observations and modeling of the FS emission reveal the burst energetics, outflow 
geometry, and the density structure of the pre-explosion environment, the self-similar hydrodynamic 
evolution of the FS is insensitive to the composition of the jet itself.

Instead, the composition (baryon content), initial Lorentz factor, and magnetization of GRB 
jets can be probed through the short-lived RS emission \citep{gk03,gt05,zk05}. The expected 
signature of RS synchrotron radiation is a bright optical flash as the shock crosses the ejecta 
(typically lasting a few tens of seconds), followed by a radio flare (typically lasting a few 
days), a phenomenon predicted to be prevalent, if not ubiquitous, in GRBs \citep{abb+99,sp99a}. 
Isolating the RS requires careful decomposition of the observed multi-frequency (radio to 
X-ray) spectral energy distribution (SED) at different epochs into FS and RS contributions 
\citep{lbz+13,pcc+14,vdhpdb+14,lab+16,alb+17,lbm+18,lab+18}.

RS emission is expected to be polarized, particularly if the jet contains large-scale ordered 
magnetic fields advected from the central engine \citep{gk03,gt05}, and thus RS polarization 
observations provide a powerful measure of ejecta magnetization \citep{mkca+13}. 
The degree of RS 
polarization is sensitive to the magnetic field anisotropy in the jet, with levels of up to 
$\approx60\%$ expected in the presence of ordered magnetic fields or $\lesssim10\%$ in the case of 
tangled fields \citep{gra03,gk03,lpb03}. 
The polarization angle is predicted to remain stable in jets with large-scale magnetic fields 
\citep{lcg+04}, or vary randomly with time if the field is produced locally by plasma or 
magnetohydrodynamic (MHD)
instabilities \citep{gw99}. {Thus measurements of polarization degree and position angle, 
and of the evolution of these quantities with time should provide diagnostics for the magnetic 
field structure in GRB jets.}

Polarized RS emission in the radio or millimeter band has not been detected to date due to 
sensitivity and response time limitations, with the best limits of $\lesssim 7\%$ (linear) 
and $\lesssim 9\%$ (circular) for likely RS emission in GRB\,991216 at 8.46\,GHz, $1.5$\,d after the 
burst \citep{gt05}, and $<3.9\%$ (linear) and $<2.7\%$ (circular) at 1.5~days after the burst for 
the strong RS observed in GRB\,130427A \citep{vdhpdb+14}. 
However, RS emission, although visible for up to $\sim$ a week in 
the cm-band, is often self-absorbed at these frequencies \citep{lbz+13,lab+16,lbm+18,lab+18}; as 
self-absorption suppresses intrinsic polarization \citep{tin08}, this could potentially explain the 
cm-band upper limits. In contrast, RS emission is expected to be optically thin in the mm-band; 
however, the limited sensitivity and response time of mm-band facilities has precluded such a 
measurement to date.

Here, we present Atacama Large Millimeter/submillimeter Array (ALMA) Band 3 (97.5~GHz)
full Stokes observations of GRB~190114C, beginning at 
2.2~hours after the burst and lasting for 3~hours, together with NSF's Karl G.~Jansky Very
Large Array (VLA) observations spanning 
4.7--6.3~hours after the burst. 
Our data reveal the first polarization detection\footnote{We note that a manuscript reporting 
detection of polarized mm-band emission in GRB~171205A at $\approx5.2$~days after the burst 
appeared at \url{https://arxiv.org/abs/1904.08111} two days after our article was posted on arXiv.} 
at radio or millimeter frequencies. By combining 
the radio, millimeter, optical, and X-ray observations, we demonstrate that the mm-band 
flux is dominated by a reverse shock, which allows us to constrain the magnetic field geometry in 
the outflow powering this burst. We assume $\Omega_m=0.31$, $\Omega_{\lambda}=0.69$, and 
$H_0=68$\,km\,s$^{-1}$\,Mpc$^{-1}$. All times are relative to the \Swift\ trigger time and in the 
observer frame, unless otherwise indicated. 

\section{GRB properties and observations}
\subsection{Gamma-ray, X-ray, and optical}
GRB~190114C was discovered by the Burst Alert Telescope \citep[BAT;][]{bbc+05} on the 
\textit{Neil Gehrels Swift Observatory} \citep[\Swift;][]{gcg+04} on 2019 January 14 at 
20:57:03\,UT \citep{gcn23688}. The GRB was also detected by \Konus\ in the 30~keV to 20 MeV band, 
which observed decaying emission until $\approx320$~s after the trigger \citep{gcn23737}, by the 
\Fermi\ Gamma-ray Burst Monitor \citep[GBM;][]{mlb+09} with $T_{90}\approx116$~s, and by the 
\Fermi\ Large Area Telescope \citep{gcn23707,gcn23709}. 
In a historic first, {high-energy} emission from this burst was also detected by the twin 
Major Atmospheric Gamma Imaging Cherenkov (MAGIC) telescopes at $\gtrsim300$~GeV, starting 50~s 
after the BAT trigger \citep{gcn23701}. 

The optical afterglow was discovered by the \Swift\ UV/Optical 
Telescope \citep[UVOT;][]{rkm+05} starting 73~s after the BAT trigger \citep{gcn23688}. 
Spectroscopic observations with the ALFOSC instrument on the Nordic Optical Telescope (NOT) 
beginning $\approx29$~minutes after the BAT trigger yielded a redshift of $z=0.42$ 
\citep{gcn23695}, which was further refined with X-shooter spectroscopy at the European Southern 
Observatory's Very Large Telescope to $z=0.4245$ \citep{gcn23710}.

The \Swift X-ray Telescope \citep[XRT;][]{bhn+05} began observing GRB~190114C 64~s after the BAT 
trigger. 
We use the photon index for WT- and PC-mode observations listed on the \Swift\ website, together 
with the corresponding unabsorbed counts-to-flux conversion rate to convert the 0.3--10\,keV count 
rate light curve\footnote{Obtained from the \Swift\ website at 
\url{http://www.swift.ac.uk/xrt_curves/883832} and re-binned to a minimum signal-to-noise ratio per 
bin of 10.} to flux density at 1\,keV. 
We performed photometry on the UVOT data at $6\times10^{-3}$~days with a 3.5\arcsec aperture 
(including aperture corrections) using standard techniques \citep{pbp+08}. 
We further include observations of the 
afterglow reported in GCN circulars, in particular, the NOT observations at $2\times10^{-2}$~days 
\citep{gcn23695} and the GROND observations at $\approx0.16$~days \citep{gcn23702}. 

\subsection{Radio: VLA}
We observed the afterglow using the Karl G. Jansky Very Large Array (VLA) starting 4.7~hours 
($\approx0.2$~days) after the burst through program 18A-088 (PI: Laskar). In our first epoch, we 
obtained a full sequence of observations spanning 5--38\,GHz. We used 3C48 as the flux density and 
bandpass calibrator and J0402-3147 as the complex gain calibrator. 
We carried out data reduction 
with CASA \citep{mws+07} using the \texttt{pwkit} package \citep{wcnr17}. The highest frequency (K 
and Ka band) observations exhibited significant post-calibration residuals, which we remedied 
using phase-only self-calibration. We are continuing observations of the afterglow in the cm band 
at the time of writing, and defer a detailed analysis of the cm-band properties of this event at 
$\gtrsim1$~day to a future work. We list the results of our VLA observations in Table 
\ref{tab:stokesI}.

\begin{deluxetable}{ccccc}
 \tabletypesize{\footnotesize}
 \tablecolumns{5}
 \tablecaption{Radio and millimeter Stokes $I$ observations of GRB~190114C}
 \tablehead{   
   \colhead{Telescope} &
   \colhead{Frequency} &
   \colhead{Time} &
   \colhead{Flux density} &
   \colhead{Uncertainty}\\
   \colhead{} &
   \colhead{(GHz)} &
   \colhead{(days)} &
   \colhead{(mJy)} &
   \colhead{($\mu$Jy)}
   }
 \startdata 
ALMA & $97.5$ &$0.0995$ & $11.5$ & $21.8$ \\
ALMA & $97.5$ &$0.107$ & $11.1$ & $21.2$ \\
ALMA & $97.5$ &$0.115$ & $10.7$ & $25.7$ \\
ALMA & $97.5$ &$0.125$ & $10.2$ & $31.4$ \\
ALMA & $97.5$ &$0.129$ & $10.1$ & $60.8$ \\
ALMA & $97.5$ &$0.140$ & $9.58$ & $20.8$ \\
ALMA & $97.5$ &$0.146$ & $9.26$ & $35.5$ \\
ALMA & $97.5$ &$0.154$ & $8.60$ & $29.6$ \\
ALMA & $97.5$ &$0.161$ & $8.24$ & $22.2$ \\
ALMA & $97.5$ &$0.168$ & $8.05$ & $25.5$ \\
ALMA & $97.5$ &$0.188$ & $7.53$ & $23.0$ \\
ALMA & $97.5$ &$0.196$ & $7.27$ & $23.1$ \\
ALMA & $97.5$ &$0.203$ & $7.04$ & $23.7$ \\
ALMA & $97.5$ &$0.213$ & $7.00$ & $28.8$ \\
ALMA & $97.5$ &$0.217$ & $6.87$ & $56.7$ \\
VLA & $37.0$ & $0.197$ & $3.95$ & $39.0$ \\
VLA & $30.0$ & $0.197$ & $3.28$ & $32.0$ \\
VLA & $24.5$ & $0.219$ & $2.66$ & $23.0$ \\
VLA & $19.2$ & $0.219$ & $1.96$ & $18.0$ \\
VLA & $16.0$ & $0.236$ & $1.52$ & $24.4$ \\
VLA & $13.5$ & $0.236$ & $1.22$ & $29.9$ \\
VLA & $11.0$ & $0.249$ & $0.838$  & $19.0$ \\
VLA & $8.55$ & $0.249$ & $0.607$  & $17.3$ \\
VLA & $7.10$ & $0.261$ & $0.397$  & $19.3$ \\
VLA & $5.00$ & $0.261$ & $0.118$  & $32.0$ 
\enddata
\label{tab:stokesI}
\end{deluxetable}

\subsection{ALMA polarization observations}
We obtained ALMA observations of GRB\,190114C beginning 2.2 hours after the burst through 
program 2018.1.01405.T (PI: Laskar) in full linear polarization mode in Band 3, with two 4 GHz-wide 
base-bands centered at 91.5 and 103.5 GHz, respectively. Weather conditions were excellent during 
the observation. The calibration sources were selected by ALMA, employing J0423-012 as flux 
density, bandpass, and polarization leakage calibrator, and J0348-274 as 
complex gain calibrator. The gain calibrator-source cycle time was $\approx12$ minutes, with 10.5 
minutes on source, 30 seconds on the gain calibrator, and the remaining time used for slewing 
between the two. The scheduling block was repeated three times in succession in order to achieve 
sufficient parallactic angle coverage to simultaneously derive the instrumental polarization and 
the Stokes parameters of the leakage calibrator, with parallactic angle coverage on the 
leakage calibrator spanning $\approx90^\circ$.

\subsubsection{ALMA Data analysis}
We processed the ALMA data using CASA \citep{mws+07}, employing standard techniques \citep{nnp+16}. In 
summary, following bandpass calibration, we computed the complex gain solutions on the 
polarization calibrator. We used these solutions to estimate the intrinsic Stokes parameters of the 
polarization calibrator, followed by the cross-hand delays, the XY-phase offset, and the 
calibrator's intrinsic polarization. We resolved the phase ambiguities in the Stokes parameters of 
the calibrator using the estimates derived from the gain calibration, and revised the gain solutions 
on the polarization calibrator. The ratio of the parallel hand (XX/YY) gains is uniform and 
within $\approx2\%$ of unity for all antennas after polarization calibration, while the rms gain 
ratio is uniform across antennas at the $\approx1.2\%$ level. 
The leakage (D-terms) were found to be at the $\approx1\%$ level for individual antennas, as 
expected for the ALMA 12m array \citep{nnp+16}. 

We used flux density values of $(4.15\pm0.08)$ mJy at 91.5\,GHz and $(3.89\pm0.06)$ mJy at 103.5 
GHz for J0423-012 from the ALMA calibrator catalog, to which we fit a power law model to fix the 
flux density scale for each channel. We subsequently calibrated the remainder of the dataset using 
standard interferometric techniques (flux density and gain), and generated Stokes $IQUV$ images of 
the calibrators and the target, as well as an image of the total linear polarization, 
$P=\sqrt{Q^2+U^2}$. 

\begin{figure}
  \centering
  \includegraphics[width=\columnwidth]{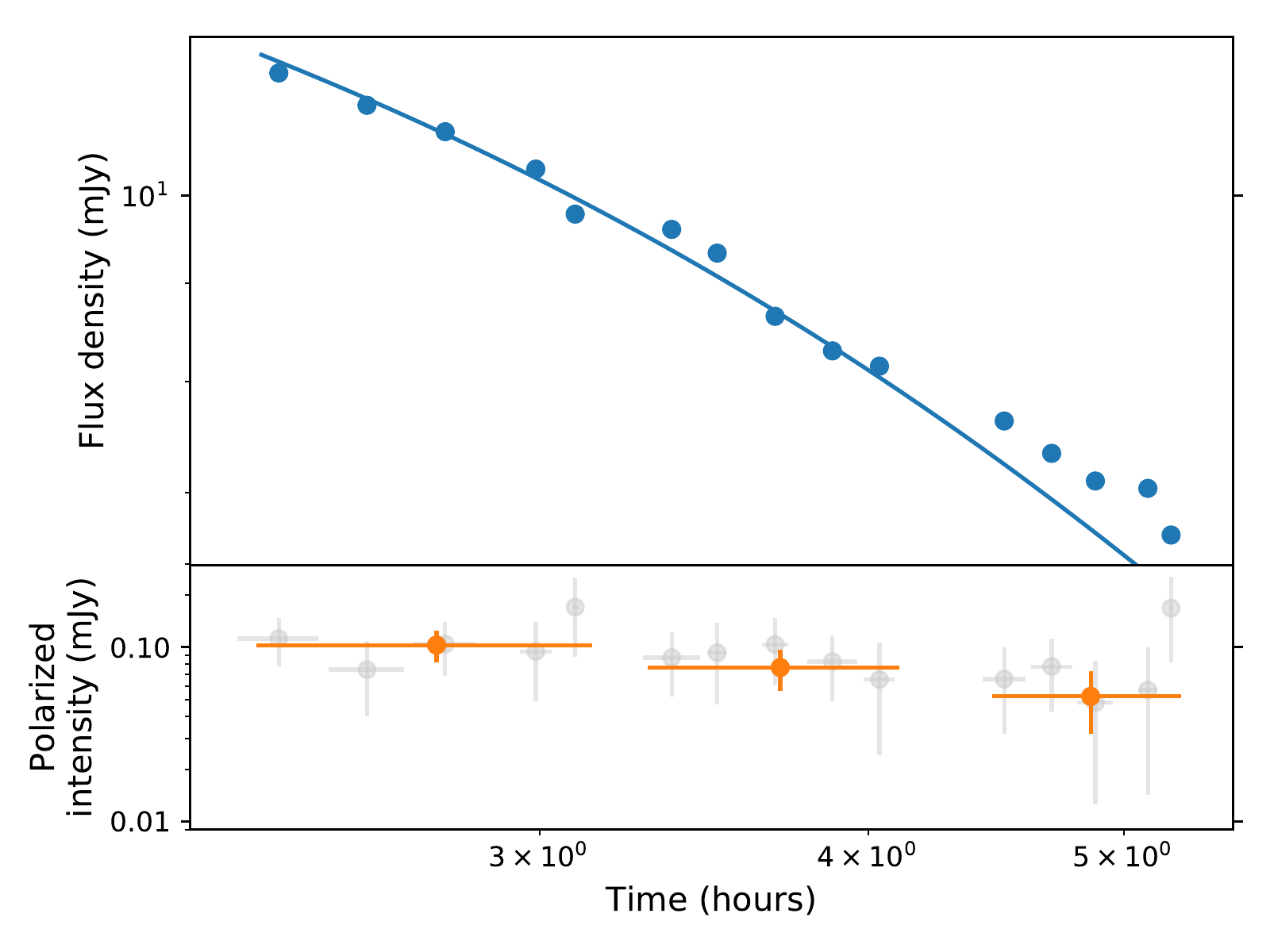}
  \caption{Total intensity (Stokes $I$) and reverse shock model (top panel; Section \ref{text:rs}) 
  and linear polarized intensity obtained from a Monte Carlo analysis (grey, and binned orange 
points, lower panel) for our ALMA observations at 97.5~GHz spanning 2.2 -- 5.2 hours (134 -- 
313 minutes) after the burst. The afterglow fades by $\approx40\%$ during these first three hours 
of 
observation, while the polarized intensity drops by $\approx 50\%$.}
\label{fig:grb_pollc}
\end{figure}

The mm-band afterglow is clearly detected in Stokes $I$, with a signal-to-noise of $\approx580$, 
allowing us to divide the source data set into individual scans. We fit for the flux density of the 
source in the image plane using \texttt{imfit}. The derived flux density values are listed in Table 
\ref{tab:stokesI}. The mm afterglow fades by $\approx40\%$ between 2.2--5.2 hours after the burst 
(Fig.~\ref{fig:grb_pollc}; top panel).

\begin{figure*}
  \centering
   \includegraphics[width=\textwidth]{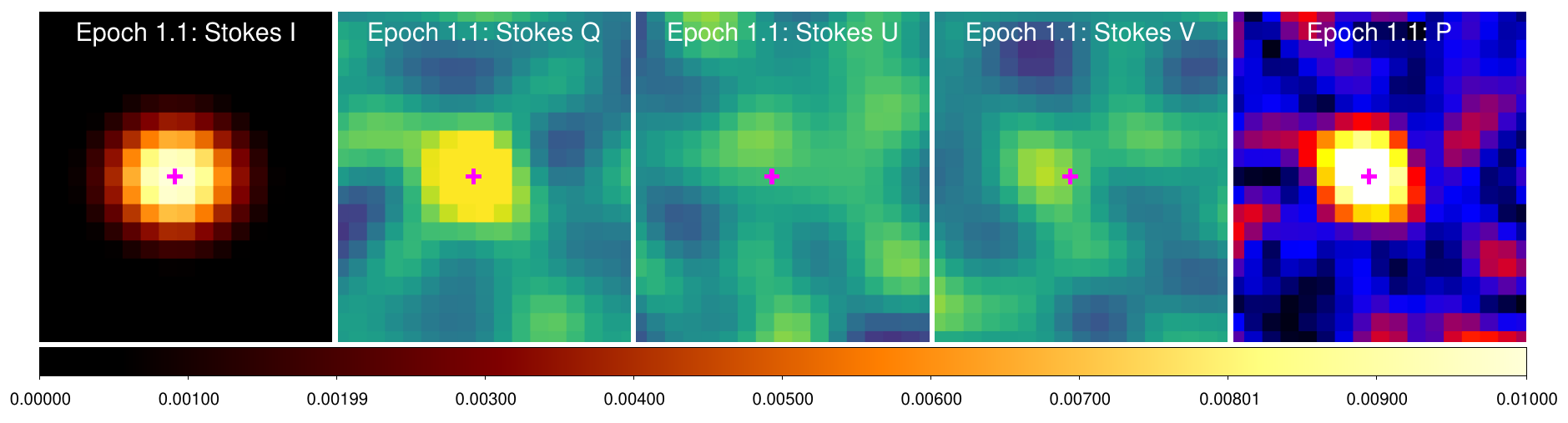}  \\
   \includegraphics[width=\textwidth]{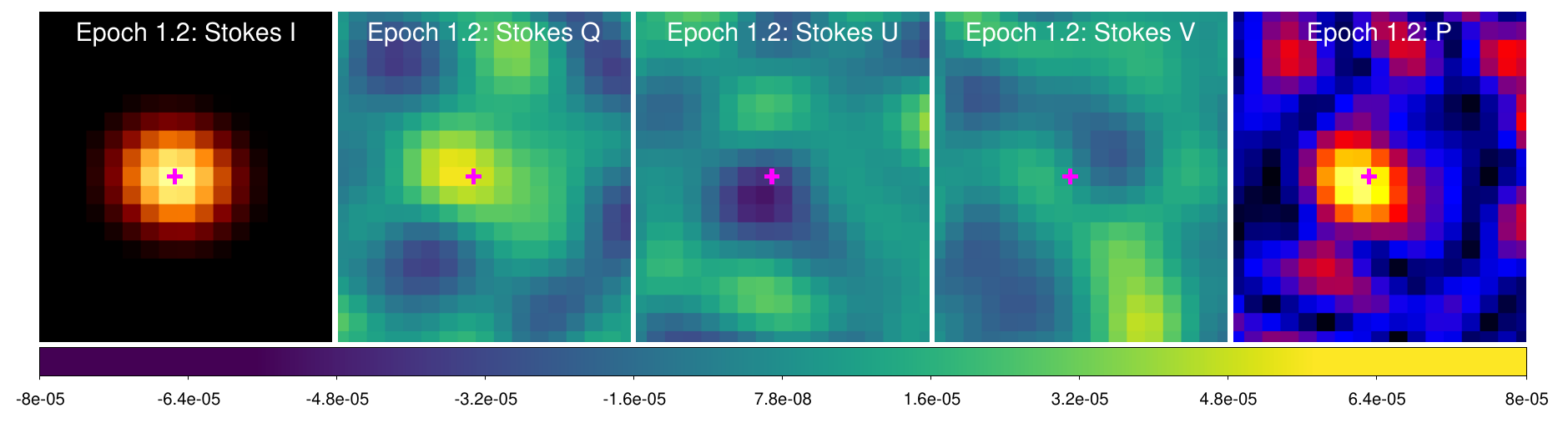}  \\
   \includegraphics[width=\textwidth]{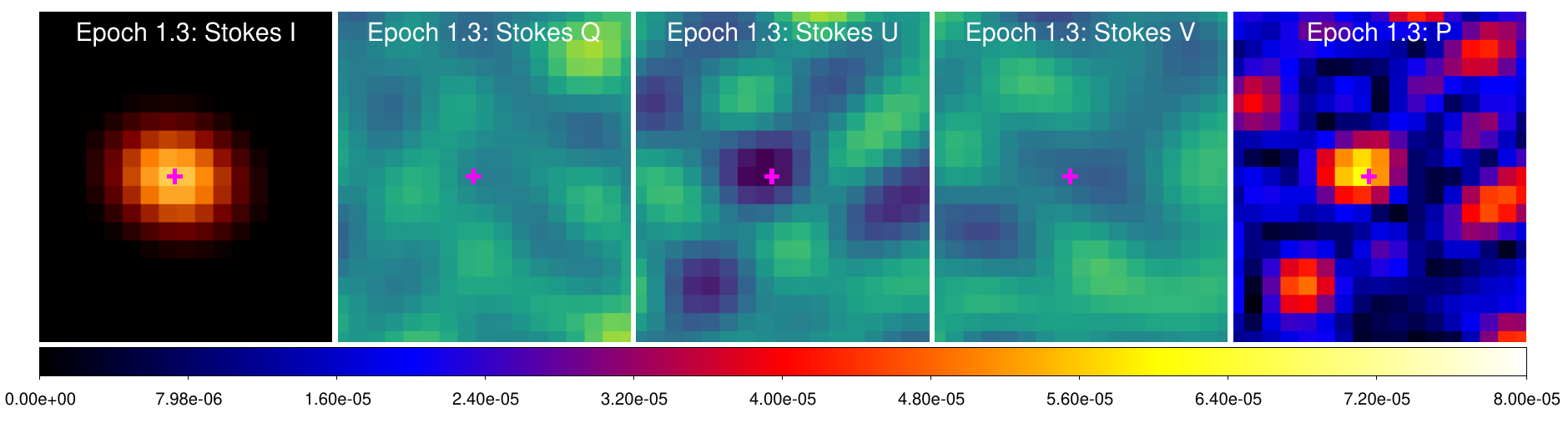}
  \caption{Stokes $IQUV$ and linear polarized intensity, $P = \sqrt{Q^2+U^2}$ images of the mm-band 
emission at mean times of 2.74 hours (top), 3.70 hours (center), and 4.86 hours (bottom) after the 
burst. All images in the same column have the same color bar and scaling parameters. The three 
color 
bars from top to bottom provide the scales for the Stokes $I$, Stokes $QUV$, and $P$ images, 
respectively. The images demonstrate the fading Stokes $I$ emission, as well as the rotation of the 
polarization angle (evolving $U/Q$ ratio).}
\label{fig:grbimages}
\end{figure*}

\begin{figure*}
  \centering
  \begin{tabular}{cc}
  \includegraphics[width=\columnwidth]{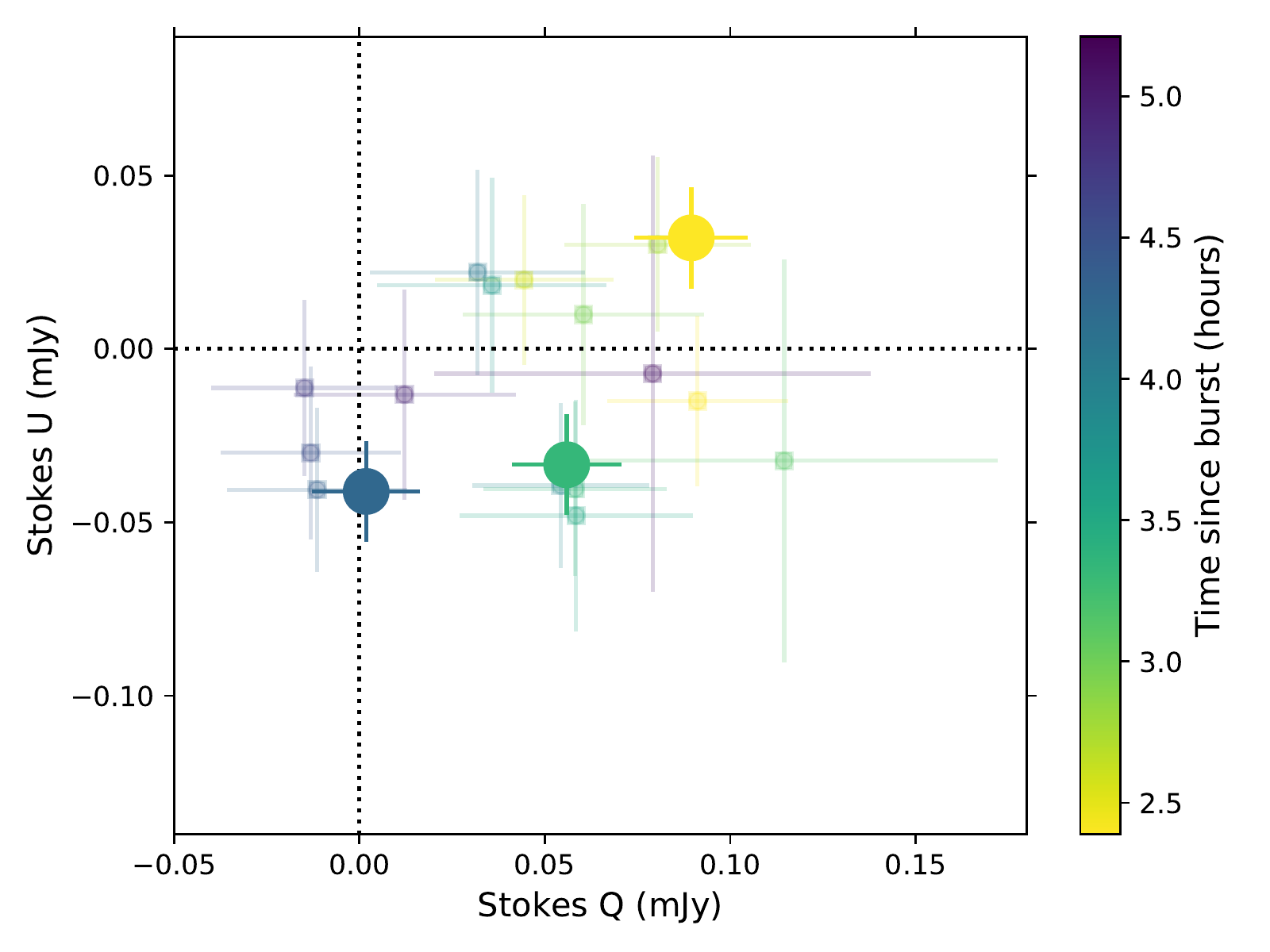} &  
  \includegraphics[width=\columnwidth]{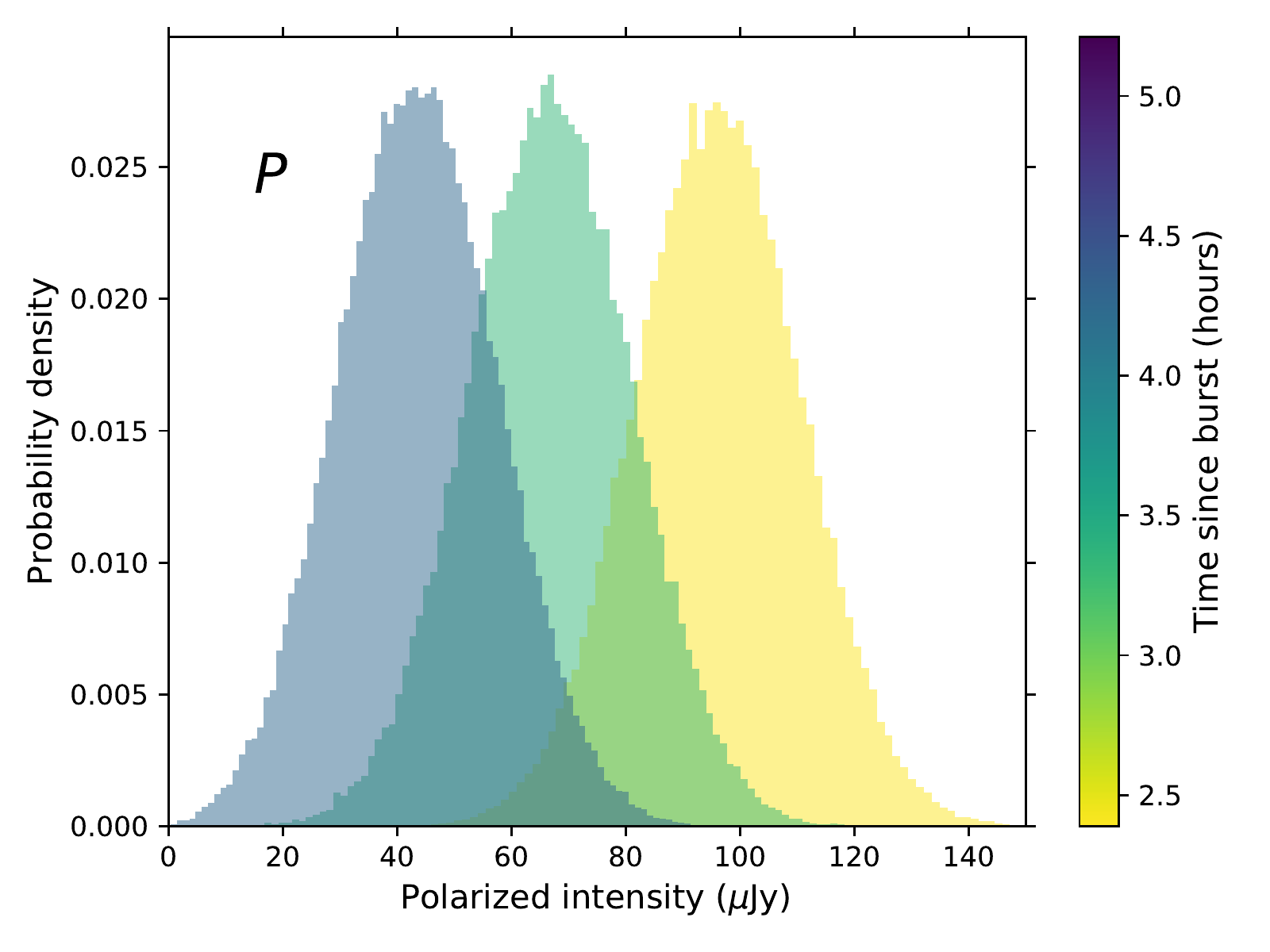}
  \end{tabular}
  \caption{\textit{Left panel:} Stokes $Q$ vs.~Stokes $U$ for our ALMA observations of 
GRB\,190114C at 2.2--5.2 hours after the burst, colored by the time of observation from earliest 
(yellow) to last (violet). Small squares correspond to individual scans on the target and have 
large uncertainties, while large circles correspond to measurements from images made with combined 
$uv$-data from each of the three executions of the scheduling block. The degree of linear 
polarization is represented by the magnitude of the vector from the origin to the (Q,U) 
measurement, 
while the polarization position angle ($\chi = \frac{1}{2}\tan^{-1}\frac{U}{Q}$) is equal to 
one-half the angle subtended by that vector and the $x$-axis. The plot has the same linear scale on 
both axes, with the origin 
displaced. 
\textit{Right panel:} probability density for the total polarized intensity at 
mean times of 2.74 hours (yellow), 3.70 hours (green), and 4.86 hours (blue) after the burst, 
generated by sampling from the distributions of the individual measured Stokes $Q$ and $U$ values 
for these three epochs, and assuming Gaussian errors. The polarized intensity decreases with time.}
\label{fig:grb_qu}
\end{figure*}

\subsubsection{Measurement of polarization and validation against potential instrumental effects}
\label{text:pol}
The $P$ image of GRB\,190114C reveals a point source {with flux density $61\pm14~\mu$Jy 
(undebiased)} at a position consistent with the position in the Stokes $I$ image 
(Fig.~\ref{fig:grbimages}). 
The rms noise level in the Stokes $QUV$ images is $\approx10\,\mu$Jy. We split the $uv$ data into 
the three individual runs of the scheduling block, and re-imaged the target. The detection in the 
first $P$ image is {$6.6\sigma$ (statistical)}, and the polarized intensity declines by 
$\approx50\%$ over the course of the observation (Fig.~\ref{fig:grb_pollc}; bottom panel). The limit 
on Stokes $V$ is 30~$\mu$Jy, corresponding to a formal $3\sigma$ limit on circular polarization of 
$\Pi_{\rm V} < 0.3\%$ (statistical only) relative to the mean Stokes $I$; however, the $1\sigma$ 
systematic circular polarization calibration uncertainty is $\approx0.6\%$.

\begin{figure*}
  \centering  
  \begin{tabular}{cc}
  \includegraphics[width=\columnwidth]{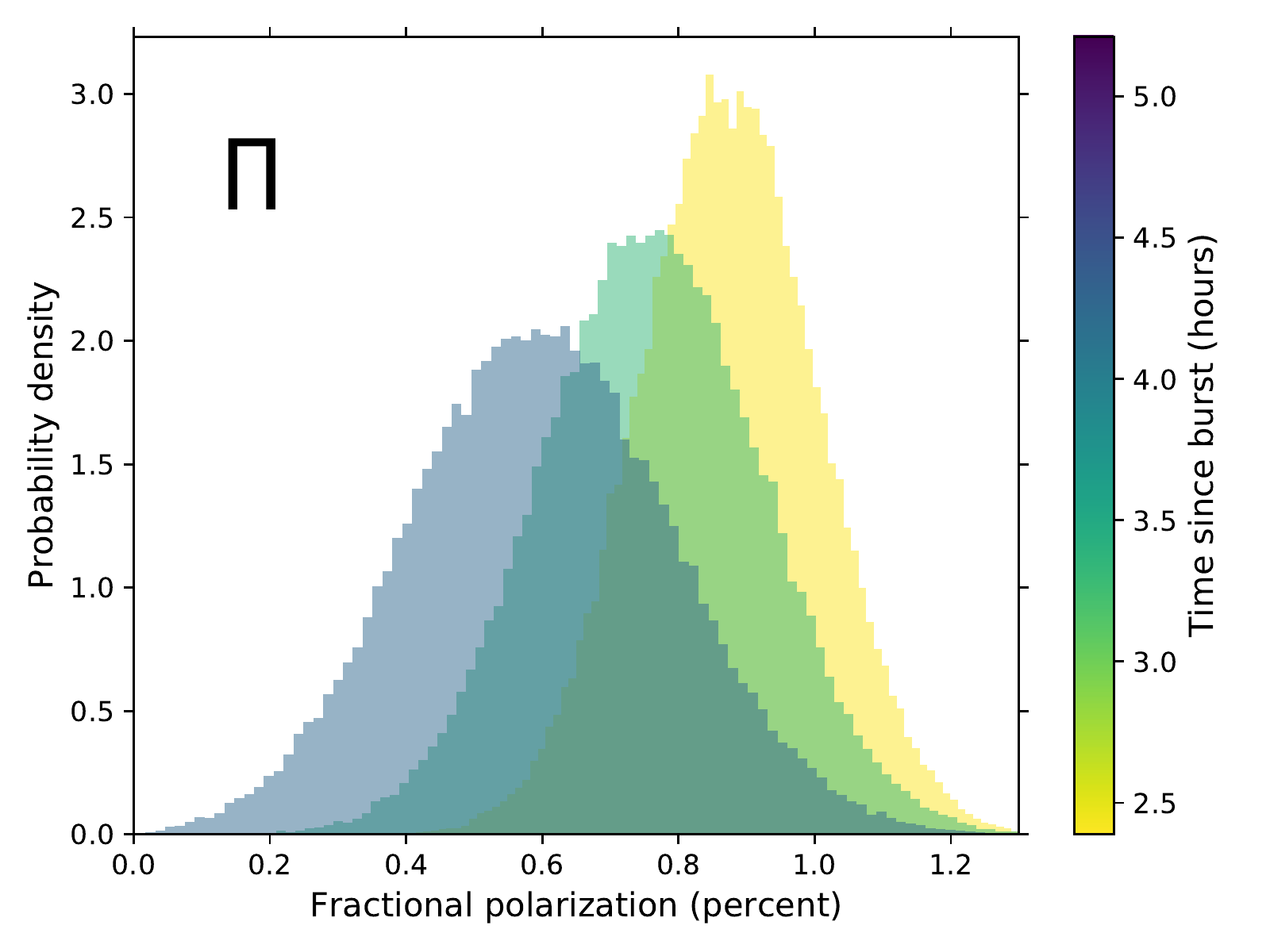} &
  \includegraphics[width=\columnwidth]{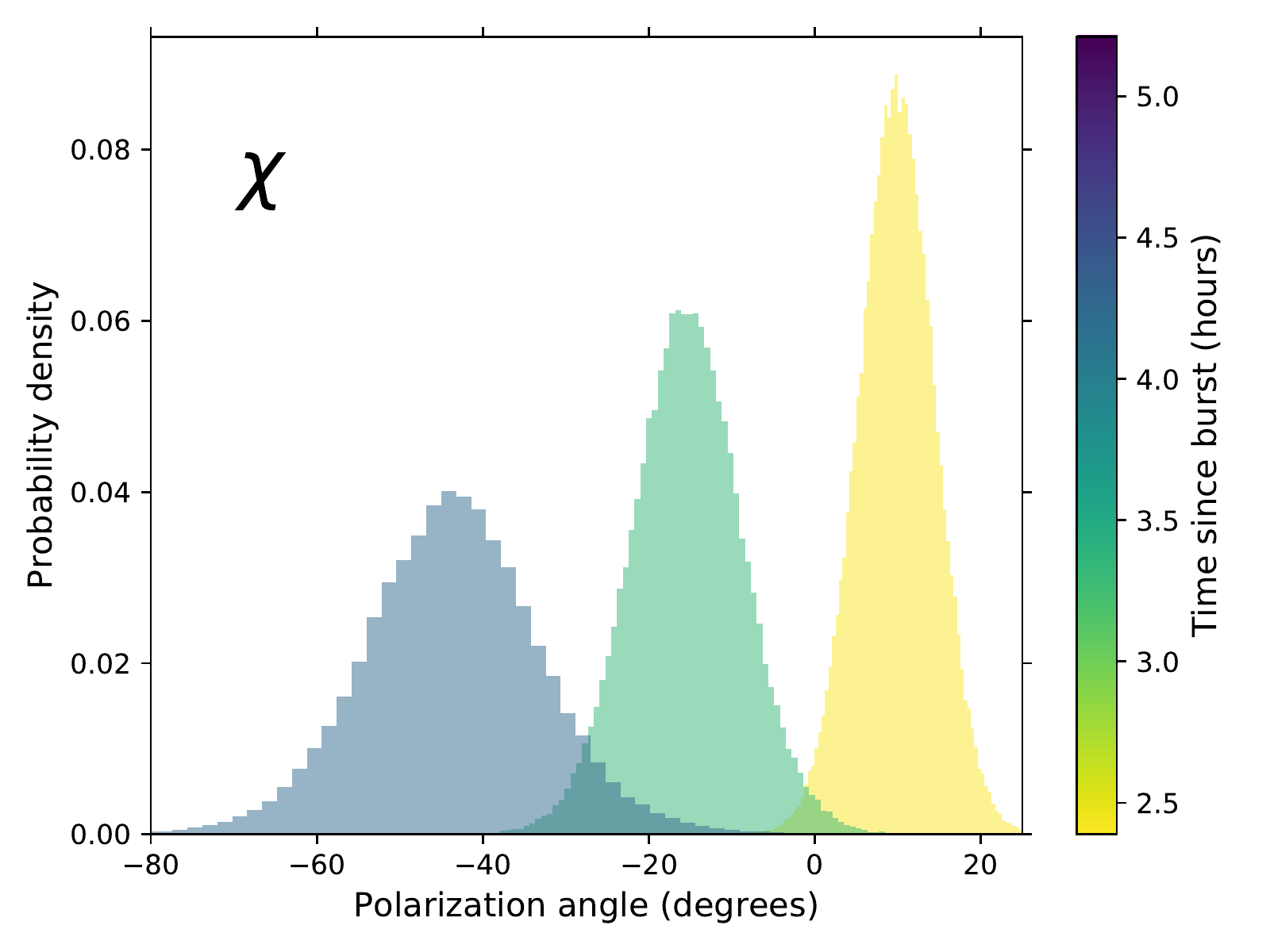} \\
  \includegraphics[width=\columnwidth]{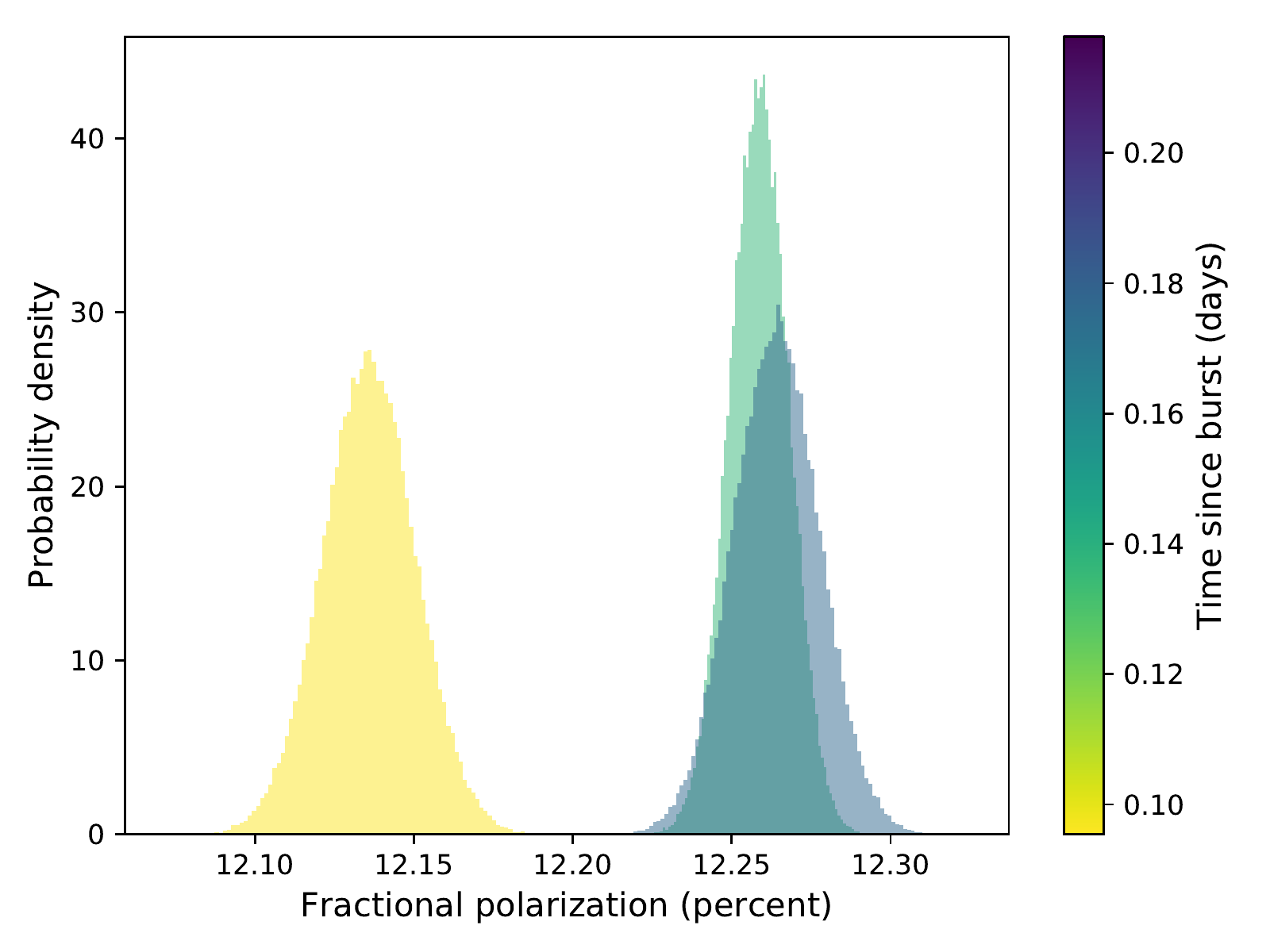} &
  \includegraphics[width=\columnwidth]{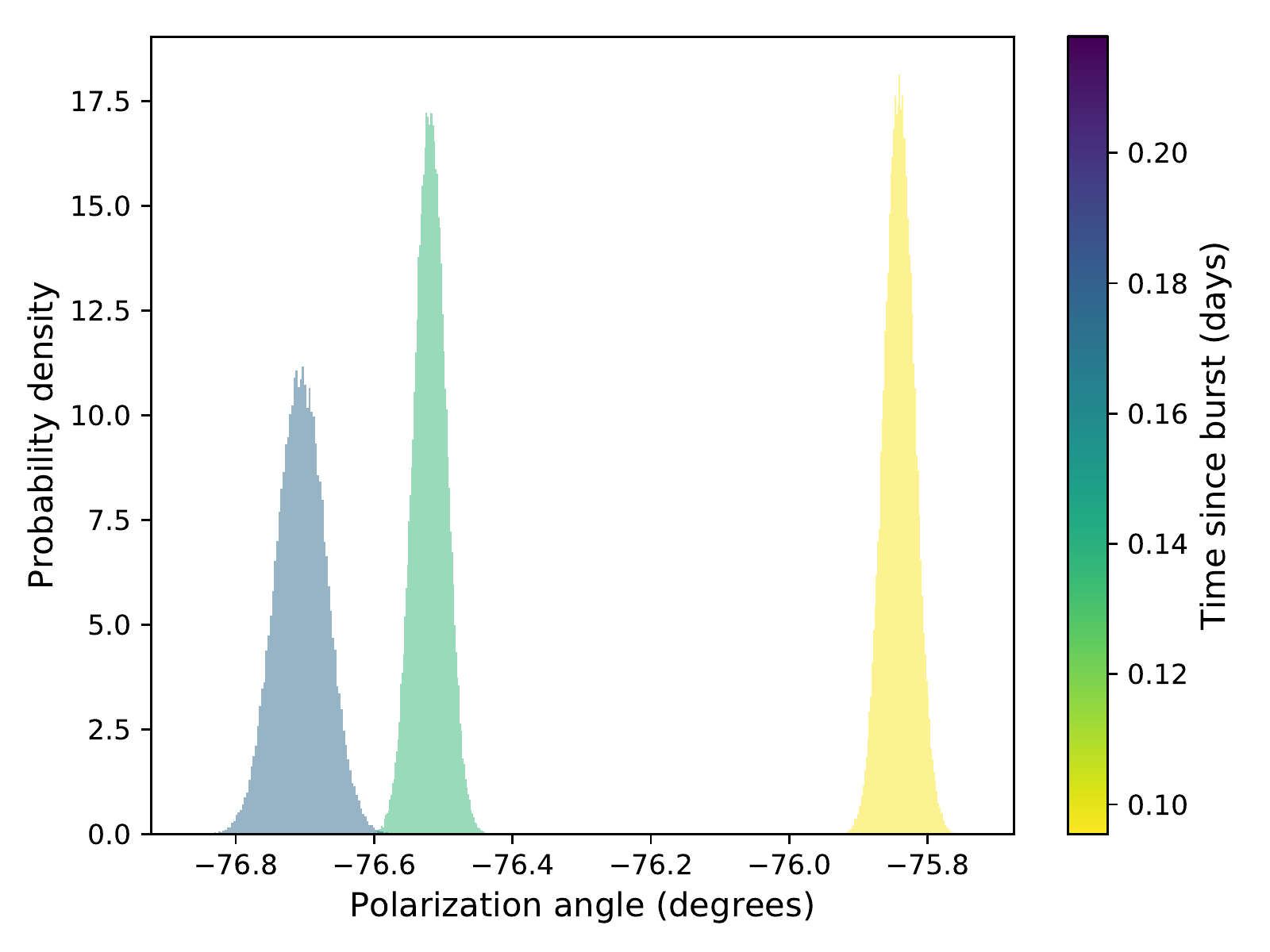} 

  \end{tabular}
  \caption{\textit{Top row:} probability density for the fractional linear polarization ($\Pi 
= P/I$, left panel) and  polarization angle ($\chi$, right panel) at mean times of 2.74 hours 
(yellow), 3.70 hours (green), and 4.86 hours (blue) after the burst, generated by sampling from the 
probability density of the total polarized intensity (Figure \ref{fig:grb_qu}) and the total 
intensity in these three epochs (assuming Gaussian errors for the latter). The fractional 
polarization decreases with time, while the polarization angle rotates by 
$\delta\chi=54\pm13$~degrees. 
\textit{Bottom row:} Same as top row but for the complex gain calibrator, which was not used in 
the polarization calibration. The fractional 
polarization is stable at $\lesssim0.15\%$ and the polarization angle within $\lesssim1^{\circ}$, 
conforming to the specifications for linear polarization observations in ALMA Cycle 6. }
  \label{fig:grb_pfpa}
\end{figure*}

We plot the values of Stokes $Q$ and $U$, measured by fixing the position and beam parameters using 
the Stokes $I$ image, in Fig.~\ref{fig:grb_qu}. A rotation in the plane of polarization is apparent 
from the Stokes $QU$ images.
As images of $P$ are biased for faint sources, we do not measure $P$ from images of 
polarized intensity, but rather from the measured $QU$ values directly using a Monte Carlo method. 
We generate $10^5$ random realizations from the individual $Q$ and $U$ measurements and calculate 
$P=\sqrt{Q^2+U^2}$, the polarization angle, $\chi = \frac{1}{2}\tan^{-1}\frac{U}{Q}$, and the 
fractional linear polarization, $\Pi = P/I$. For the latter, we incorporate the uncertainty in the 
measurement of Stokes $I$. We plot the derived distributions of $P$, $\chi$, and $\Pi$ in Figures 
\ref{fig:grb_qu} and \ref{fig:grb_pfpa}, and list the median and standard deviations {of the 
distributions} in Table~\ref{tab:pol}. 

On applying the polarization calibration to the gain calibrator J0348-274, we find that drift 
in the linear polarized intensity of the calibrator is $\lesssim0.15\%$, while its measured 
polarization angle is stable at the $\lesssim 1\%$ level over the course of the 3-hour observation 
(Fig.~\ref{fig:grb_pfpa}; bottom panel). Both values are within the specifications of ALMA Cycle 6 
polarization observations. 

One possible manifestation of any errors arising from leakage calibration is a scattering of flux 
density in the Stokes $QU$ images away from the phase center. We check this by imaging the gain 
calibrator, which appears as a point source; observed secondary peaks in both Stokes $Q$ and $U$ 
images  are $\lesssim0.5\%$ of the peak flux, consistent with noise.
We also imaged the upper and lower base-bands separately for both the flux density calibrator, 
phase calibrator, and GRB\,190114C. The polarization properties of both calibrators and of 
GRB\,190114C are consistent between the two base-bands and thus stable across ALMA Band 3. 

As linear polarization observations are a non-standard mode for ALMA, the data were also calibrated 
and imaged by a data analyst (Erica Keller) at ALMA before delivery. We compared the results of our 
reduction with those from ALMA, and also by imaging the calibrated measurement set provided by the 
Observatory. All three sets of images yield results consistent within measurement uncertainty. 
These 
tests indicate that the detection of linearly polarization in GRB\,190114C is unlikely to arise 
from 
a calibration artifact.

\begin{deluxetable*}{rrrrrrrrrrr}
 \tabletypesize{\footnotesize}
 \tablecolumns{11}
  \tablecaption{ALMA Band 3 (97.5 GHz) Polarization Measurements of GRB~190114C}
  \tablehead{   
   \colhead{Time} &
   \colhead{$Q$} &
   \colhead{$\sigma_Q$} &
   \colhead{$U$} &
   \colhead{$\sigma_U$} &
   \colhead{$P$} &
   \colhead{$\sigma_P$} &
   \colhead{$\chi$} &
   \colhead{$\sigma_{\chi}$} &
   \colhead{$\Pi$} &
   \colhead{$\sigma_{\Pi}$} \\
   \colhead{(days)} &
   \colhead{($\mu$Jy)} &
   \colhead{($\mu$Jy)} &
   \colhead{($\mu$Jy)} &
   \colhead{($\mu$Jy)} &
   \colhead{($\mu$Jy)} &
   \colhead{($\mu$Jy)} &
   \colhead{(deg)} &
   \colhead{(deg)} &
   \colhead{(\%)} &
   \colhead{(\%)}       
    }
 \startdata 
0.114 & 89.5 & 14.6 &  32.0 & 15.3 & 96.3 & 14.6 &   9.8 &  4.6 & 0.87 & 0.13 \\
0.154 & 55.9 & 14.5 & -33.4 & 14.8 & 66.7 & 14.4 & -15.4 &  6.6 & 0.76 & 0.16 \\
0.202 & 1.84 & 14.5 & -41.1 & 14.6 & 43.7 & 14.0 & -43.7 & 11.7 & 0.60 & 0.19
 \enddata
 \label{tab:pol}
\end{deluxetable*}

\section{Basic considerations}
As the focus of this Letter is on the ALMA polarization observations, we defer a discussion of the 
full multi-wavelength modeling to a future work (T.~Laskar et al.~in preparation). To provide 
context for the polarization detection, here we consider the basic properties of the afterglow at 
$\lesssim0.3$~days, during the time of the ALMA observations. We interpret this under the standard 
synchrotron framework \citep{spn98,gs02}, for a given isotropic equivalent kinetic energy, 
$\EKiso$ and circumburst density parameter $n_0$ (for a constant-density environment) and $\Astar$ 
(for a wind-like environment with density, $\rho\propto\,R^{-2}$). We assume the radiation is 
produced by non-thermal electrons accelerated to a power-law distribution with energy index $p$, 
with a fraction $\epse$ of the post-shock internal energy given to relativistic electrons and a 
fraction $\epsb$ to magnetic fields. In this model, the observed SED is 
characterized by power laws connected at spectral breaks: the synchrotron self-absorption frequency 
(\nua), the characteristic synchrotron frequency (\numax), and the cooling frequency (\nuc), and is 
completely specified by the location of these break frequencies and the overall flux density 
normalization (\fnumax).

\subsection{Optical and X-rays: circumburst density profile}
The spectral index\footnote{We use the convention $f_{\nu}\propto t^{\alpha}\nu^{\beta}$ 
throughout.} between the GROND $g'$ and $K$ bands, when corrected for Galactic extinction, 
is $\beta_{\rm NIR-opt}=-2.4\pm0.2$, indicating that extinction is present. The $r'$-band light 
curve decays as $\alpha_{\rm r}=-0.69\pm0.02$ between $3\times10^{-2}$~days and 0.3~days, while the 
X-ray decay rate over this period is $\alpha_{\rm X} = -1.27\pm0.02$, indicating that the optical 
and X-rays are on different power-law segments of the synchrotron spectrum. In the slow cooling 
regime with $\numax<\nuopt<\nuc<\nux$, we expect $\delta\alpha\equiv|\alpha_{\rm opt}-\alpha_{\rm 
X}|=0.25$, which is inconsistent with the measured $\delta\alpha = 0.58\pm0.03$. The only other 
means for the optical light curve to decay slower than the X-rays is if the system is fast cooling 
with $\nuopt < \nuc < \numax < \nux$ and the circumburst density profile is a wind-like 
environment. 
In this regime, we expect $\alpha_{\rm opt}\approx-2/3$, which is 
consistent with the observed $r'$-band light curve over this period. The shallow optical light 
curve 
also places a lower limit on the jet break time, $\tjet\gtrsim0.3$~days.

The steep X-ray light curve with $\alpha_{\rm X}\approx-1.3$ in fast cooling implies 
$\nuc,\numax<\nux$, which suggests $p\approx2.36$. The observed X-ray spectral index over this 
period is $\beta_{\rm X}=-0.81\pm0.14$. Whereas this is inconsistent with a predicted slope of 
$\beta_{\rm X} \approx-1.2$, it is consistent with a spectral slope of 
$\beta\sim-(p/2-1/4)\sim-0.93$ when Klein-Nishina (KN) corrections are taken into account 
\citep{nas09}. {We note that a similar discrepancy in the X-ray spectral index of 
GRB~161219B was also attributable to KN corrections \citep{lab+18}}. We leave a detailed 
exploration of KN corrections to further work. In 
summary, the optical and X-ray light curves until 0.3~days are consistent with FS emission in 
wind-like environment with $p\approx2.36$ and $\tjet\gtrsim0.3$~days.

\subsection{Radio and millimeter: RS}
\label{text:rs}
\begin{figure*}
  \centering
  \begin{tabular}{cc}
  \includegraphics[width=\columnwidth]{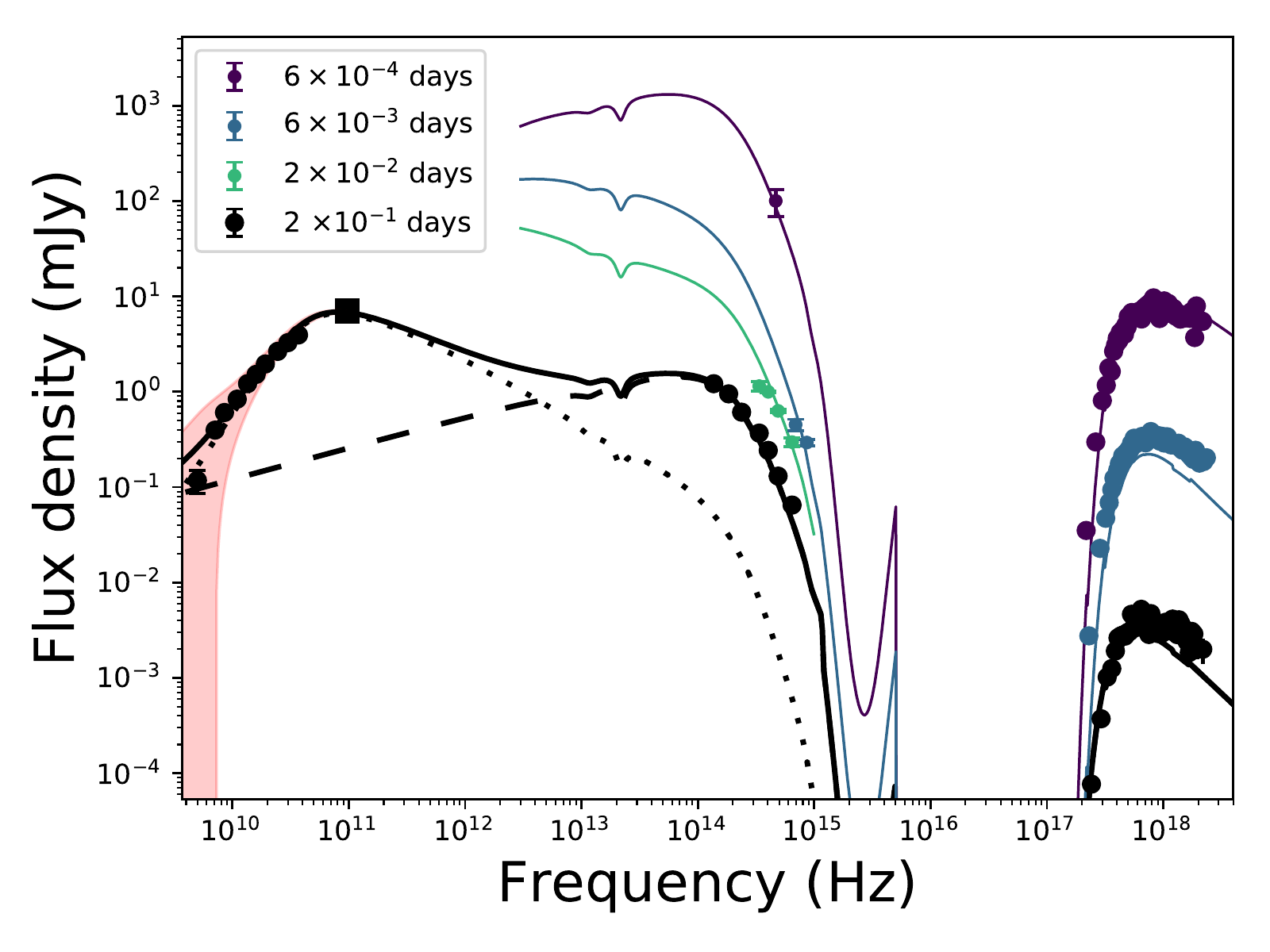}  &  
  \includegraphics[width=\columnwidth]{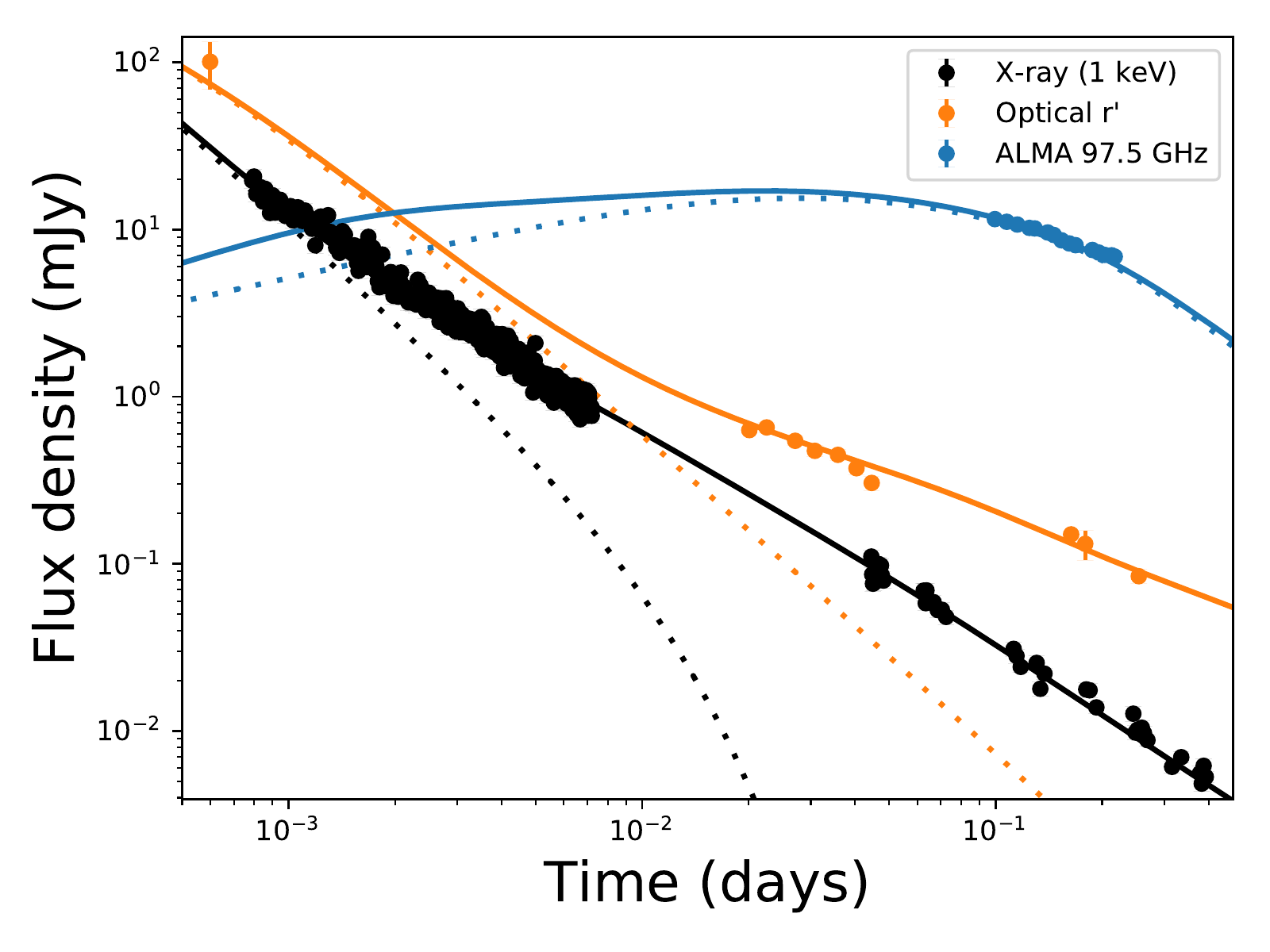}  
  \end{tabular}
  \caption{\textit{Left panel:} 
Spectral energy distributions at $6\times10^{-4}$~days (MASTER; \citealt{gcn23690}), 
$6\times10^{-3}$~days (\Swift/UVOT), $2\times10^{-2}$~days (NOT; \citealt{gcn23695}) and 0.2~days 
(VLA, ALMA, and GROND; \citealt{gcn23702}) after the burst, with an afterglow model (lines) 
decomposed at 0.2~days into forward shock (dashed) and reverse shock (dotted) components. The RS 
model employs $\nuar\approx40$~GHz, $\numr\approx70$~GHz, $\nucr\approx4\times10^{15}$~Hz, and 
$\fnumr\approx14$~mJy at 0.2~days. The red shaded region indicates the expected variability due to 
scintillation in the radio. The model explains the radio to X-ray SED, the X-ray light curve, and 
the optical light curve before 0.2~days. The Compton $Y\approx20$ for this model is high, and the 
discrepancy in the X-rays above $\approx10^{18}$~Hz may arise from the Klein-Nishina correction.
\textit{Right panel:} X-ray (1 keV), optical $r'/R/Rc$-band, and ALMA 97.5~GHz light curves of 
GRB~190114C from the first MASTER detection at $\approx6\times10^{-4}$ days to $\approx0.3$~days, 
together with the same afterglow model as the left panel, with the RS contribution indicated (dotted 
lines).
}
\label{fig:rssed}
\end{figure*}

The radio SED at 0.2~days comprising the VLA cm-band and ALMA mm-band data 
{(Fig.~\ref{fig:rssed})} can be fit with a broken power law model, transitioning from 
$\beta=2$ (fixed) to $\beta=0.3\pm0.2$ at $\nu_{\rm break}=24\pm4$ GHz. {In addition, the 
mean Stokes $I$ intra-band spectral index between the two ALMA base-bands at 91.5~GHz and 103.5~GHz 
is $\approx-0.4$, implying that the mm-band emission is optically thin at this time.} 
The optical to mm-band spectral index of $\beta_{\rm mm-opt}=-0.24\pm0.01$ between the GROND 
$K$-band observation and the ALMA detection at 0.16~days is inconsistent with a single power-law 
extrapolation from the optical\footnote{We note that extinction correction at optical $K$-band is 
expected to be modest. Explaining the declining mm-NIR SED as due to extincted FS emission would 
require $A_K\approx4.5$~mag (or $A_V\approx35$~mag for a Small Magellanic Cloud extinction curve), 
which would completely extinguish the UV/optical emission.}. This shallow slope cannot be caused 
by the location of $\numf$ between the radio and optical bands\footnote{The subscript `f' refers to 
the FS.} because all light curves at 
$\nuaf<\nu<\numf$ should be flat in the wind model (or rising in the interstellar medium (ISM) model), while the ALMA 
light curve is declining over this period. 
Thus, the radio and mm-band emission arises 
from a separate component than that responsible for the X-ray and optical emission. We note that a 
similar radio-to-X-ray spectral index of $\beta_{\rm radio,opt} \approx -0.25$ in the case of 
GRB~130427A indicated the presence of an RS in that system \citep{lbz+13}.
The early optical $r'$-band light curve declines as $\alpha_{\rm opt}=-1.4\pm0.1$ 
between the MASTER observation at $\approx6\times10^{-4}$~days\footnote{While the MASTER 
observation is calibrated to R band, the difference between $r'$ and $R$ bands is negligible for 
this argument.} and the NOT observation at $\approx2\times10^{-2}$~days, flattening to 
$\alpha-0.69\pm0.02$ between the NOT observation and the GROND observation at $0.16$~days 
{(Fig.~\ref{fig:rssed})}. The {steep optical light curve at $\lesssim 
2\times10^{-2}$~days} can also not be explained as FS emission. 

We find that propagating the excess emission component dominating the radio and mm-band data at 
$\approx0.2$~days {earlier, using the RS light curve evolution from \cite{zwd05} and the 
SED shape from \cite{lbz+13},} can explain the optical observations at 
$<0.2$~days, provided 
$\fnumax\propto t^{-0.9}$ and $\numax\propto t^{-1.4}$ for this component (Fig.~\ref{fig:rssed}). 
This matches a Newtonian RS with\footnote{The Lorentz factor of the reverse-shocked ejecta, 
$\Gamma_{\rm ej}\propto R^{-g}$.} $g\sim3$, which is higher than expected for the wind environment 
but not unprecedented \citep{lbz+13,pcc+14,lab+16,lab+18}. The parameters for the FS that match the 
X-ray and optical light curves at $\lesssim0.3$~days are $p\approx2.36$, $\epse\approx0.9$, 
$\epsb\approx6\times10^{-3}$, $\Astar\approx1.5\times10^{-2}$, $\EKiso\approx7\times10^{52}$~erg, 
and $A_{\rm V}\approx2.2$~mag. For these parameters, the FS is fast cooling until 
$\approx0.2$~days, with the spectral ordering $\nuaf<\nu_{\rm radio}<\nucf\approx\nuopt<\numf<\nux$ 
for the forward shock at $10^{-2}$~days. However, we note that we do not locate $\nuaf$ and thus 
the model parameters are subject to some degeneracies (possibly explaining the high value of 
$\epse$). We defer a more complete analysis of the FS and the joint RS--FS 
dynamics to future work.

\section{Discussion}
We now derive constraints on the magnetic field structure in the jet using our polarization 
measurement. 
The low level of measured linear polarization in the mm-band,\footnote{The observed low degree of 
linear polarization is unlikely to result from Faraday depolarization, as the latter is strongly 
suppressed at these frequencies \citep{gt05}. Furthermore, we find no evidence of increased 
polarization upon decreasing our observing bandwidth by splitting the data 
into the two base-bands (Section~\ref{text:pol}).}, $\Pi\sim0.6-0.9\%$, rules 
out an ordered transverse magnetic field ($B_{\rm ord}$) in the ejecta with an angular coherence 
length $\theta_B\gtrsim 1/\Gamma$, where $\Gamma$ is the Lorentz factor of the emitting region, as 
such a field would produce a polarization of several tens of percent \citep{gra03,gk03,lpb03}. We 
now consider scenarios where the received radiation is a superposition of distinct emission 
components in regions comprising a transverse ordered field ($B_{\rm ord}$) on the one hand, and a 
random ($B_{\rm rand}$) magnetic field \citep{gk03} on the other.
Such a scenario might correspond to co-located field components such as a shocked ISM with an 
ordered upstream field compressed at the FS and a random shock-generated $B_{\rm rand}$, or to the 
superposition of emission from two distinct regions, e.g. a dominant $B_{\rm ord}$ in the shocked 
ejecta and a dominant $B_{\rm rand}$ in the shocked ISM. 
In such scenarios, $\Pi$ and $\chi$ depend on the ratio of the intensities of synchrotron radiation 
due to the two magnetic field components, $I_{\rm ord}/I_{\rm rand}\approx\langle B_{\rm 
ord}^2/B_{\rm rand}^2\rangle$, and thus can vary with time \citep{gk03}. However, {the 
significant change in $\chi$ we measure} would require comparable polarized intensities from the 
two components, with a ratio varying on the dynamical time. This is not easy to realize at $\ll 
\tjet$, and where the 97.5~GHz light curve is dominated by RS emission, and thus such scenarios are 
disfavored.

Next, we consider a model where the observed polarization is the sum of emission from 
intrinsically polarized but mutually incoherent patches, each with a magnetic field ordered over a 
typical angular scale, $\theta_{\rm B}$ \citep{gk03,no04,gt05}. 
In this model the visible region, $\theta\sim1/\Gamma_{\rm ej}$ around the line of sight gradually 
increases as the jet decelerates. The number of patches contributing to the observed emission is 
given by $N\sim(\Gamma_{\rm ej}\theta_{\rm B})^{-2}$. In general, the ejecta lags behind the forward 
shock 
and $\Gamma_{\rm ej}\lesssim\Gamma_{\rm sh}$; however, a Newtonian RS does not 
significantly decelerate the ejecta \citep{kob00}. For $g\approx3$ and $k=2$, we have $\Gamma_{\rm 
ej}/\Gamma_{\rm sh}\propto (t/t_{\rm dec})^{-[g-(3-k)/2]/[(4-k)(2g+1)]}\propto (t/t_{\rm 
dec})^{-5/28}$ \citep{gt05}. Taking $\tdec\approx T_{90}=116$~s from 
\Fermi/GBM\footnote{The UVOT light curve is definitely declining by $\tdec=566$~s \citep{gcn23725}. 
Taking $\tdec$ equal to this upper limit only increases $\Gamma_{\rm ej}$ by $\approx30\%$.}, 
$\Gamma_{\rm ej}/\Gamma_{\rm sh}\approx0.5$ at the time of our mm-band polarization measurement. At 
this time, the {Lorentz factor of the fluid shocked by the FS, 
$\Gamma_{\rm sh}=3.7\left[\frac{E_{\rm K,iso,52}(1+z)}{\Astar t_{\rm 
days}}\right]^{1/4}\approx30$} \citep{gs02}, so that $\Gamma_{\rm ej}\approx15$.

The maximum degree of polarization, $\Pi_0=(1-\beta)/(5/3-\beta)$, where $\beta$ is the spectral 
index \citep{gt05}. Since the ALMA band is near the peak of the SED 
(Fig.~\ref{fig:rssed}), we take $\beta\sim0$, yielding $\Pi_0\sim0.6$. The observed 
polarization is a random walk of $N$ steps in the $QU$ plane, with $\Pi \sim \Pi_0/\sqrt{N} \sim 
\Pi_0\Gamma_{\rm ej}\theta_B$, which implies $\theta_{\rm B}\sim \Pi/(\Gamma_{\rm 
ej}\Pi_{0})\approx10^{-3}$. The uncertainty on this estimate from the signal-to-noise of the 
measurement of $\Pi$ is $\approx15\%$; however, larger systematic uncertainties arise from the 
approximations used in the RS dynamics as well as the stochastic nature of the 2D random 
walk.

In this model, the polarization angle is expected to vary randomly over the dynamical time scale as 
new patches enter the visible region. The mm-band light curve spans a factor of $\approx2.2$ in 
time. During this period, $\Gamma_{\rm sh}$ declines from $\approx34$ to $\approx28$ from our 
afterglow model and $\Gamma_{\rm ej}$ declines from $\approx16$ to $\approx11$. Assuming 
$\theta_{\rm B}$ remains constant, the number of emitting patches increases by a factor of 
$\approx2$ over this period, which may be sufficient to change the average $\chi$ as we observe. 
Whereas we expect fluctuations in $\Pi$ over this period, our measurements do not have sufficient 
signal-to-noise to resolve such variations (Fig.~\ref{fig:grb_qu}).

Finally, we note that the gradual change observed in {$\chi$ rules out any globally 
axisymmetric magnetic field configuration, regardless of our viewing angle and of the jet's exact 
axisymmetric angular structure; for example: (i) a global toroidal magnetic field  
\citep{lcg+04,gt05}; and (ii) an axisymmetric jet} viewed from an angle $\theta_{\rm obs}>0$ from 
its symmetry axis together with a shock-produced random magnetic field $B_{\rm rand}$ that is 
symmetric around the local shock normal (tangled in three dimensions on angular scales $\ll 
1/\Gamma$, with some non-negligible degree of anisotropy, as a locally isotropic field would produce 
no net polarization), as in this case the direction of polarization is expected to remain constant 
well before the jet break time \tjet\ \citep{gl99,sar99,gk03}.

\section{Conclusions}
We present the first detection {and measurement of the temporal evolution} of linearly 
polarized emission in the radio/millimeter afterglow of a GRB, and validate that our measurement 
does not arise from a calibration artifact. 
Our detection constitutes the first measurement of a polarized RS signature at radio or 
millimeter frequencies.
The degree of linear polarization decreases from $\Pi=(0.87\pm0.13)\%$ to $\Pi=(0.60\pm0.19)\%$ 
from 2.2 to 5.2~hours after the burst, and the polarization position angle rotates from 
$\chi=(10\pm5)^\circ$ to $\chi=(-44\pm12)^\circ$ over this period. The smooth variation in $\chi$ 
rules out {axisymmetric models such as a global toroidal field in the GRB jet.} If the 
emission 
arises from small patches of coherent magnetization, then the size of these regions is constrained 
to $\theta_{\rm B}\approx10^{-3}$~radian. 
Future work on GRB~190114C that evaluates the degeneracies in the FS parameters and 
compares the derived properties of the forward and RSs to infer the dynamics of the jet, 
may refine these parameters. ALMA polarimetric observations of a sample of GRBs will reveal 
whether sub-percent levels of polarization are ubiquitous, thus constraining global jet models.

\acknowledgements
We thank Mark Lacy and Robert Laing for helpful discussions, Erica Keller at ALMA for providing 
the calibrated measurement sets for a verification of the science results, {and the 
anonymous referee for their feedback.} 
This Letter makes use of the following ALMA data: ADS/JAO.ALMA\#2018.1.01405.T. ALMA is a 
partnership of ESO (representing its member states), NSF (USA) and NINS (Japan), together with NRC 
(Canada), NSC and ASIAA (Taiwan), and KASI (Republic of Korea), in cooperation with the Republic of 
Chile. The Joint ALMA Observatory is operated by ESO, AUI/NRAO and NAOJ. VLA observations for this 
study were obtained via project 18A-088. The National Radio Astronomy Observatory is a facility of 
the National Science Foundation operated under cooperative agreement by Associated Universities, 
Inc. KDA acknowledges support provided by NASA through the NASA Hubble Fellowship grant 
\#HST-HF2-51403.001 awarded by the Space Telescope Science Institute, which is operated by the 
Association of Universities for Research in Astronomy, Inc., for NASA, under contract NAS5-26555. 
JG and RG are supported by the Israeli Science Foundation under grant No.~719/14. The Berger 
Time-Domain Group at Harvard is supported in part by NSF under grant AST-1714498 and by NASA under 
grant NNX15AE50G. RBD acknowledges support from the National Science Foundation under Grant 
1816694. This work makes use of data supplied by the UK Swift Science Data Centre at the University 
of Leicester and of data obtained through the High Energy Astrophysics Science Archive Research 
Center On-line Service, provided by the NASA/Goddard Space Flight Center. 

\bibliographystyle{apj}
\bibliography{grb_alpha,gcn}
\end{document}